# Electrically steered conduction topologies and period-doubling phase dynamics in VO$_2$

Siyuan Huang[1,2], Shuaishuai Sun[1], Yin Shi[1], Wentao Wang[3], Chunhui Zhu[4], Huanfang Tian[1,2], Huaixin Yang[1,2], Jun Li[*1], Jianqi Li[*1,2,5]

[1]Beijing National Laboratory for Condensed Matter Physics, Institute of Physics, Chinese Academy of Sciences, Beijing, 100190, China

[2]School of Physical Sciences, University of Chinese Academy of Sciences, Beijing, 100049, China

[3]Laboratory for Ultrafast Transient Facility, Chongqing University, Chongqing, 401331, China

[4]Hebei Key Laboratory of Photophysics Research and Application, Hebei Normal University, Shijiazhuang 050024, China

[5] Songshan Lake Materials Laboratory, Dongguan 523808, China

**Abstract**

The insulator-to-metal transition (IMT) in strongly correlated materials, such as vanadium dioxide (VO$_2$), offers a transformative platform for next-generation adaptive electronics and neuromorphic computing[1]. However, harnessing this non-equilibrium phase transition for deterministic device operation is fundamentally hindered by the inability to disentangle electric-field effects from Joule heating, owing to a lack of *operando* techniques capable of resolving phase dynamics at nanoscale spatial and sub-nanosecond temporal scales[2]. Here, using a newly developed electrical-pulse-pump ultrafast transmission electron microscope (E-UTEM), we directly visualize the multi-scale electro-thermo-mechanical dynamics of the IMT in suspended VO$_2$ devices. Our results reveal that electric-field-induced Poole-Frenkel (PF) emission, localized by patterned oxygen vacancies, plays a decisive role in redistributing the internal electric field to trigger a deterministic Mott transition. The extreme non-linearity of this PF effect enables the formation of dynamically reconfigurable connectivity topologies that bypass conventional thermal limits. Furthermore, we observe that the coupling of thermal and elastic energies governs a discrete domain evolution, characterized by step-wise and period-doubling configurational resets, which is a hallmark of non-equilibrium phase dynamics in constrained geometries[3]. By integrating experimental imaging with phase-field simulations, we establish a comprehensive framework for the electrically-driven IMT and predict sub-100-ps switching kinetics. These findings provide a fundamental basis for the rational design of ultrafast, low-energy functional devices through nanoscale defect and strain engineering in correlated systems.

**Introduction**

The exquisite interplay among charge, lattice, and orbital degrees of freedom in strongly correlated electron systems manifests as a rich landscape of emergent phenomena, including insulator-to-metal transition (IMT)[4], superconductivity[5], and charge density wave[6]. A defining characteristic of these materials is the coexistence of competing phases, where nanoscale domains form intricate patterns that dictate the macroscopic physical properties. Vanadium dioxide ($VO_2$) has emerged as a canonical model for exploring such physics, as it undergoes a dramatic first-order IMT coupled with a structural phase transition (SPT) from a low-temperature monoclinic (M1) phase to a high-temperature metallic rutile (R) phase[7]. Beyond its fundamental interest, the ability to deterministically manipulate these phase domains—controlling their nucleation, growth, and spatial topology[8-10]—offers a transformative pathway for developing "reconfigurable matter" and adaptive electronic components that transcend the limits of conventional silicon technology[11].

However, harnessing the electrically driven IMT for deterministic device operation remains fundamentally hindered by an enduring controversy: the inability to disentangle the contributions of direct electric-field effects from Joule heating during the switching process[12]. Despite advances in ultrafast electron diffraction[13], ultrafast spectroscopy[14] and time-resolved X-ray techniques[15], these methods either lack spatial resolution or cannot track nanoscale nucleation events, including metallic filament formation and domain wall migration in real space[16-18]. This restricted spatiotemporal access to nucleation sites, propagation trajectories, and domain boundary dynamics has precluded definitive discrimination between field- and thermally-driven mechanisms. Consequently, the underlying physics of filamentary conduction remains obscured, hindering the exploitation of non-linear phenomena such as directional percolation and topological reconfiguration for advanced logic operations.

Here, we employ operando electrical-pulse-pump ultrafast transmission electron microscopy (E-UTEM) to capture the real-space evolution of the electrically driven IMT within suspended monocrystalline $VO_2$ devices. By utilizing ultrafast dark-field (UDF) imaging, we identify a multiphysics coupling where Poole-Frenkel (PF) emission serves as the primary catalyst, localizing both current density and the internal electric field to trigger a deterministic Mott transition. Unlike conventional isotropic percolation, the exponential carrier multiplication inherent to the PF effect facilitates an abrupt, non-linear transformation of the conduction topology. We demonstrate that this pathway can be spatially "programmed" through electron-beam-induced oxygen vacancy patterning, enabling precise control over the device's functional connectivity. Furthermore, our observations reveal that the transient interplay between thermal gradients and elastic constraints actively reshapes the underlying energy landscape.

This interaction manifests as a rhythmic morphological evolution of triangular domains, characterized by continues geometric scaling and quantized-like periodic configurational resets. By resolving these non-equilibrium dynamics at the nanoscale, we not only settle the mechanistic debate surrounding field-driven phase transitions but also introduce a robust framework for designing reconfigurable logic architectures through the manipulation of domain topologies.

**Result**

***Operando* E-UTEM platform and quasi-static domain evolution under continuous current**

The experimental architecture for the electrical-pulse-pump and pulsed-electron-beam-probe E-UTEM is schematically illustrated in Fig.1a (detailed configurations are provided in Methods and Extended Data Fig. 1a). To enable *operando* visualization of the IMT, the VO$_2$ two-terminal devices were integrated onto *in-situ* electrical chips using focused ion beam (FIB) milling (see Method and Extended Data Fig. 1b for details). Our setup utilizes a digital delay generator to provide precisely timed excitation pulses, while the incident ($V_i$) and transmitted ($V_t$) voltages across the device are monitored via high-speed oscilloscopes (terminal impedance $R_s$ = 50 Ω) to track the transient device resistance (Extended Data Fig. 1d). The resulting structural and phase evolutions are probed by pulsed electron beam, synchronized to the electrical excitation at a 10 kHz repetition rate (Fig. 1b), ensuring high-fidelity capture of the non-equilibrium dynamics.

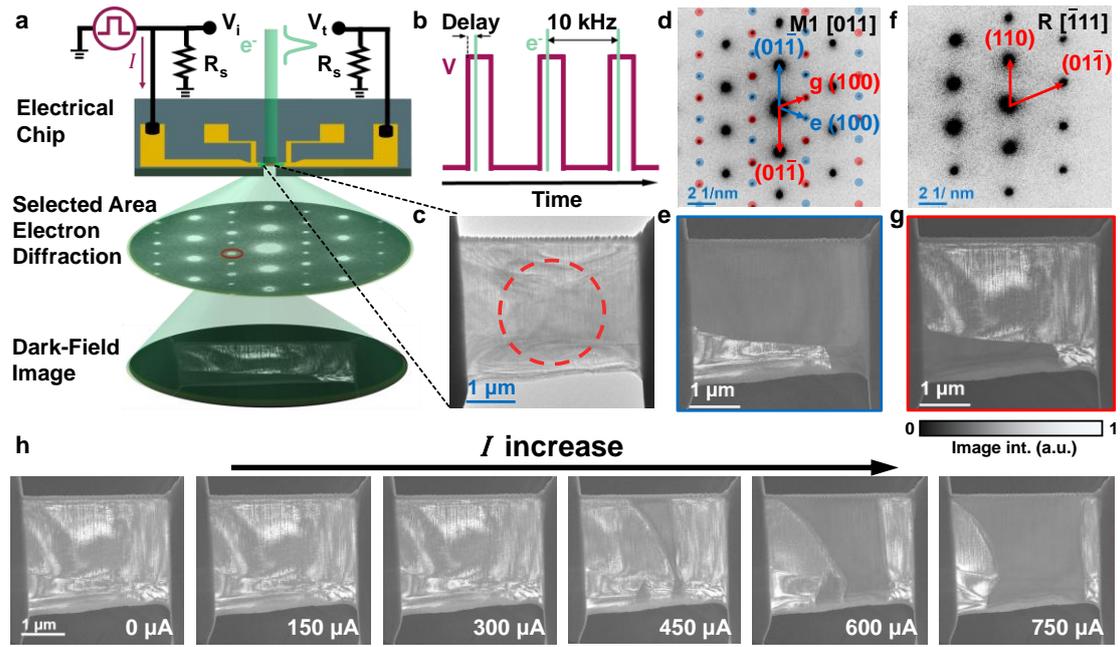

**Fig. 1. Experimental configuration and phase transition dynamics. a,** Schematic of experimental setup, integrating electrical pulse excitation with selected area electron diffraction (SAED) and dark-field (DF) imaging. The VO$_2$ specimen is suspended across the slit of an *in-situ* electrical chip. Incident ($V_i$) and transmitted ($V_t$) voltages are monitored via an oscilloscope with a terminal impedance ($R_s$) of 50 Ω. **b,** Synchronization scheme between the pulsed electron probe and 10-kHz excitation pulses, featuring a tunable temporal delay. **c,** Bright-field TEM image of the VO$_2$ device. **d,** SAED pattern obtained from the region indicated by the red circle in **c** at equilibrium (unexcited). The red and blue shading highlight the superlattice spots from two distinct monoclinic domain variants, which exhibit mirror symmetry with respect to the $(01\bar{1})_{M1}$ plane. **e,** DF image formed using the $(100)_{M1}$ spot (blue arrow in **d**). **f,** SAED pattern from the same region under 750 µA continuous-current excitation. **g,** DF image formed using the $(100)_{M1}$ spot (red arrow in d). **h,** Superimposed DF images (from **e** and **g**) illustrating the spatial distribution of phase evolution driven by a continuous current swept to 750 µA at a rate of 10 µA s$^{-1}$.

The bright-field TEM image in Fig. 1c illustrates the VO$_2$ device geometry, with the red circular region denoting the area probed by selected-area electron diffraction (SAED). In the equilibrium state, the SAED pattern (Fig. 1d) is indexed along the $[011]_{M1}$ zone axis, exhibiting two sets of diffraction spots that are mirror-symmetric with respect to the $(01\bar{1})_{M1}$ plane (highlighted in red and blue respectively). Upon driving the IMT with a 750 μA continuous current, the pattern transforms to the $[\bar{1}11]_R$ zone axis (Fig.1f). The SPT is marked by the extinction of the $(100)_{M1}$ superlattice spots originating from the Peierls distortion in the M1-phase and a concomitant intensity enhancement of the fundamental Bragg spots.

To spatially resolve these structural variants, we employed dark-field TEM (DF-TEM) imaging. By selectively filtering the mirror-symmetric $(100)_{M1}$ spots, we explicitly mapped the distribution of the corresponding M1 domains (Fig. 1e, g). Correlating the evolution of these distinct DF signals allows for a deterministic mapping of the bias-driven SPT dynamics. Surpassing the limitations of conventional bright-field microscopy, UDF imaging enables orientation-resolved tracking of individual domains and a comprehensive reconstruction of the phase-coexistence network. Crucially, the enhanced diffraction contrast resolves domain boundaries as distinctly trackable interfaces rather than obscure structural features.

The dynamics of the SPT, driven by a continuous current swept to 750 μA at a rate of 10 μA s$^{-1}$, are presented in Fig. 1h (See Supplementary Video 1 for the full evolution). The corresponding *I-V* characteristic is shown in Extended Data Fig. 1c. The SPT manifests as a sharp contrast reversal in the DF images. Notably, the transition preferentially nucleates at the geometric center of the two-terminal device before propagating bidirectionally toward the electrodes. This spatial progression is intrinsically governed by the thermal boundary conditions of the suspended specimen: the center acts as a thermal dissipation bottleneck, whereas the electrodes serve as heat sinks. Our electro-thermal simulations (See Methods and Extended Data Fig. 2) accurately reproduce this non-uniform temperature profile, revealing a peak at the sample midpoint that mirrors the experimental nucleation site. In contrast, an electrically triggered IMT driven by charge injection would typically initiate at the electrode-sample interfaces[12]. These findings unambiguously identify Joule heating as the primary thermodynamic driver for the IMT under continuous-current excitation.

**Pulse-voltage-driven phase transition dynamics**

To achieve high-efficiency and low-power operation in next-generation neuromorphic devices, a fundamental understanding the interplay between electric-field effects and Joule heating is paramount. We first investigated the SPT dynamics under a microsecond pulse regime (1 V, 3000 ns), with the spatiotemporal evolution captured via UDF imaging (Fig. 2a and Supplementary Video 2). The observed SPT progression basically reproduces the quasi-equilibrium process driven by continuous current, characterized by initial nucleation at the device center followed by lateral expansion and bidirectional propagation. The resulting transformed region assumes a characteristic "funnel" shape. Notably, the onset of the SPT exhibits a distinct latency of approximately 600 ns relative to the pulse leading edge. This temporal retardation, consistent with the delayed resistance response (Extended Data Fig. 4) and prior reports[13,15], is quantitatively captured by our electro-thermal simulations. The simulated spatial temperature profile at 1050 ns, where the 339 K isotherm closely replicates the experimental "funnel" profile (Fig. 2b), identifies this latency as the characteristic thermal induction time required for localized Joule heating to reach the transition threshold (See Extended Data Fig. 5a for spatial temperature profiles at different time).

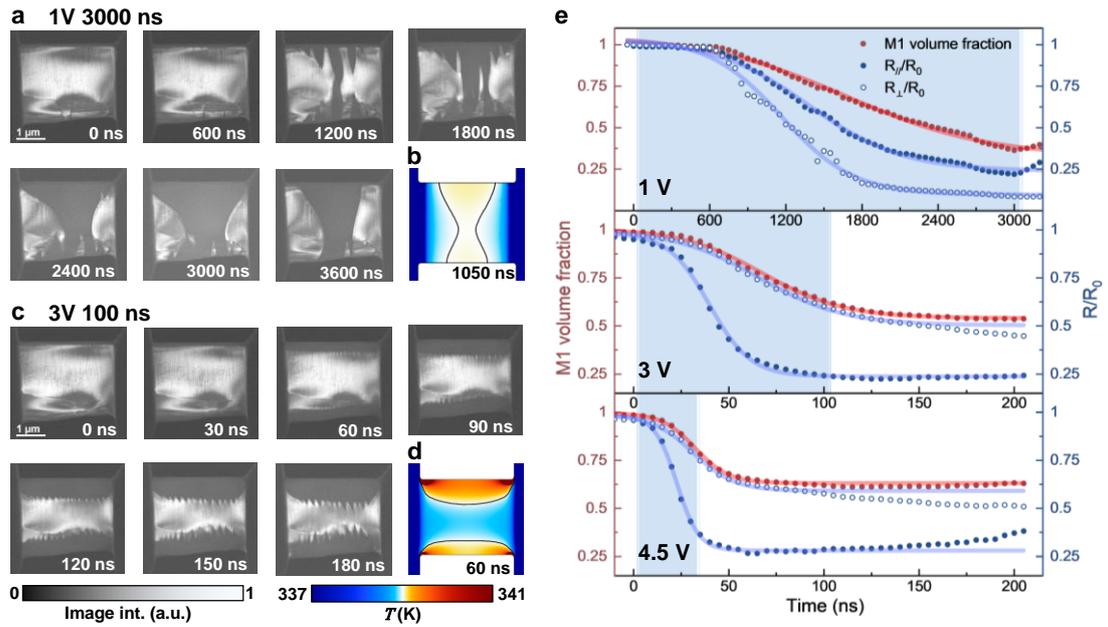

**Fig. 2. Spatiotemporal SPT dynamics and electro-thermal simulations under varying voltage pulses. a,** Time-resolved DF images capturing the SPT evolution in $VO_2$ under a microsecond electrical pulse (1 V, 3000 ns). **b,** Simulated spatial temperature profile at $t = 1050$ ns for the 1 V pulse. The black curve denotes the 339 K isotherm. **c,** Ultrafast DF images of the transient domain dynamics under a nanosecond electrical pulse (3 V, 100 ns). Note the distinct transition from the thermal "funnel" to a field-driven boundary-nucleation pattern. **d,** Simulated spatial temperature profile at $t = 60$ ns for the 3 V pulse. The black curve denotes the 339 K isotherm, highlighting the deviation from the experimental observation in **c**. **e,** Evolution of normalized resistance parallel ($R_{//}/R_0$) and perpendicular ($R_{\perp}/R_0$) to the current direction, alongside the M1 phase volume fraction, driven by different pulses amplitudes (Top to bottom: increasing voltages). Solid lines represent Boltzmann function fits (fitting parameters detailed in Extended Data Table 1). The shaded area represents the duration time of the pulse.

To access the regime of electronic excitation, which typically requires fields exceeding 1 MV m$^{-1}$ [19], we increased the pulse amplitude to 3 V while limiting the duration to 100 ns to prevent specimen damage. Under this high field condition, the SPT dynamics undergo a dramatic shift (Fig. 2c and Supplementary Video 3). In stark contrast to the center initiated thermal process observed at 1 V, the 3 V pulse triggers nucleation at the device boundaries. This leads to the formation of metallic conductive channels that bridge the electrodes and subsequently widen in the direction perpendicular to the current flow. This high-field behavior, consistently observed under 4.5 V, 30 ns pulses (Extended Data Fig. 6 and Supplementary Video 4), signifies a voltage-steered anisotropic percolation during the IMT.

To quantify the kinetics of this anisotropy percolation, we employed a resistor network model[20-22] to resolve the evolution of normalized resistance parallel ($R_{//}/R_0$) and perpendicular ($R_\perp/R_0$) to the current direction relative to the M1 volume fraction (Fig. 2e). At low voltage (1 V, Fig. 2e top panel), the system preferentially reaches the percolation threshold in the direction perpendicular to the current; consequently, $R_\perp$ undergoes a rapid collapse while $R_{//}$ closely tracks the M1 volume fraction. Conversely, high-voltage excitation steers the system toward immediate percolation along the current direction, causing $R_{//}$ to plummet abruptly and effectively decoupling macroscopic conductance from the total phase-volume fraction (Fig. 2e middle and bottom panel). These results demonstrate that the IMT in $VO_2$ transcends a simple scalar switching event, manifesting instead as a dynamically reconfigurable conduction architecture. Exploiting this voltage-tunable anisotropy to steer percolation symmetry within a single device footprint establishes a new foundation for morphologically reconfigurable logic.

The emergence of an anisotropic percolation kinetics appears counterintuitive, as the high-field switching trajectory deviates sharply from the center-initiated Joule heating model. To identify the physical origin of this discrepancy, we first performed electro-thermal simulations for the 3 V pulse regime (Fig. 2d). While the simulation predicts rapid heating at the longitudinal boundaries due to vacuum isolation and the low thermal conductivity of M1-phase $VO_2$ ($\kappa \sim 6$ W m$^{-1}$ K$^{-1}$)[23], a profound spatial mismatch remains between electro-thermal simulation and experiment (See Extended Data 5b for spatial temperature profiles at different time). Specifically, the simulated 339 K isotherms inevitably curve toward the corners near the electrodes, which act as primary heat sinks. In stark contrast, our experimental observations (Fig. 2c) reveal domain walls that remain strictly parallel to the boundary across the entire device width. This geometric divergence between the predicted thermal profile and the observed phase morphology provides compelling evidence that a purely thermomechanical description is insufficient. Instead, a localized, field-induced driving force must be active at the boundaries to steer the IMT.

**Electron energy loss spectroscopy analysis**

While the applied fields reached approximately 1 MV m$^{-1}$, they remain one to two orders of magnitude below the intrinsic threshold for a purely electric-field-driven IMT[24-26]. This discrepancy suggests that the boundary-initiated IMT must be assisted by extrinsic carriers. To pinpoint the origin of this field-induced driving force, we performed Electron Energy-Loss Spectroscopy (EELS) to probe the chemical and electronic environment at the device boundaries. STEM-HAADF and EELS mapping at the top boundary (Fig. 3a) reveal a gradient of oxygen vacancies introduced during focused ion beam (FIB) processing[27], a factor known to strongly modulate the electronic structure of $VO_2$[28-32]. Near the boundary, the V $L_{3,2}$-edge exhibits a pronounced red-shift and an increased $L_3/L_2$ white-line intensity ratio, signifying the reduction of $V^{4+}$ to $V^{3+}$ due to oxygen deficiency. This is further corroborated by the suppression of the O $K$-edge pre-peak ($t_{2g}$), a direct fingerprint of a reduced vanadium oxidation state[30,31,33]. Projections of the EELS mapping (Fig. 3a, right panel) reveal that these oxygen vacancy-rich regions extend approximately 50 nm into the specimen, serving as a reservoir of localized trap states. Similar spectral features were observed at the bottom boundary (Extended Data Fig. 7), confirming a symmetric distribution of reduced vanadium valence.

These findings point toward a Poole-Frenkel (PF) excitation mechanism[19,27], where the local electric field lowers the potential barrier for trapped electrons, facilitating their excitation into the upper Hubbard band. The electrical conductivity $\sigma$ dominated by the PF effect is described by[34,35]:

$$\sigma = e\mu n \exp\left(-\frac{e\phi_T - e(e|\vec{E}|/\pi\varepsilon_0\varepsilon)^{1/2}}{k_B T}\right) \quad (1)$$

where $e$ is the elementary charge, $\mu$ is the electron mobility, $n$ is the density of states in the conduction band, $\varepsilon_0$ is the vacuum permittivity, $\varepsilon$ is the relative permittivity of the M1 phase, $\phi_T$ is the ionization potential of oxygen vacancy, and $k_B$ is the Boltzmann constant. The PF effect orchestrate the IMT through two distinct physical modalities. The first one, as described by Equation (1) and Fig. 3b, is a synergistic electro-thermal pathway. By exciting trap-state carriers to the upper Hubbard band, the PF effect boosts local conductivity and accelerates the Joule-heating-driven IMT. In the second, the PF effect acts as a direct electronic driver. Under sufficiently intense electric fields (Fig. 3c), the density of generated carriers surpasses the critical Mott criterion, triggering a non-thermal IMT.

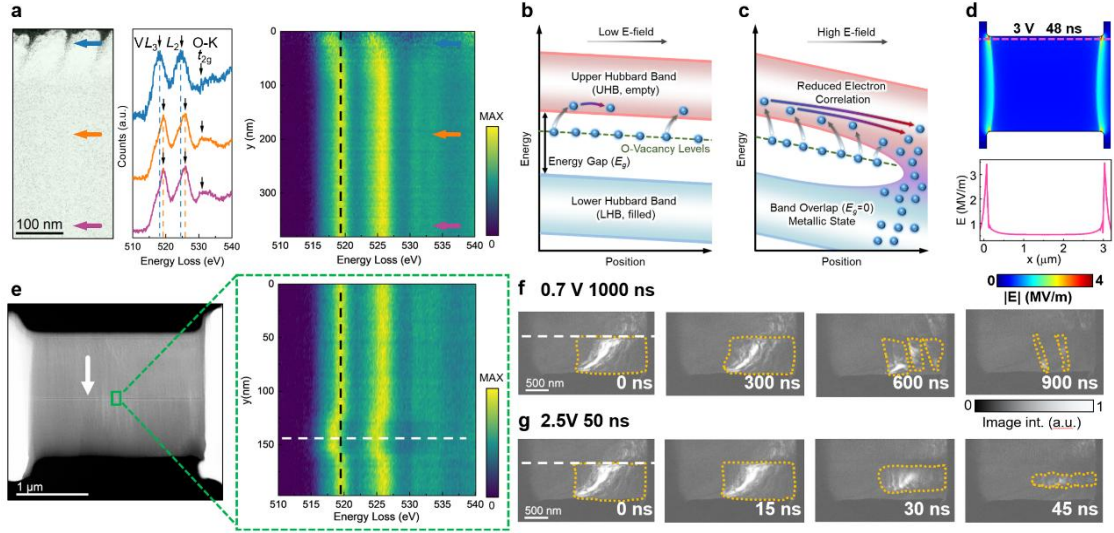

**Fig. 3. EELS characterization of oxygen vacancy-assisted IMT and deterministic control of conduction pathways. a,** EELS analysis at the top device boundary. Left: STEM-HAADF image of the EELS mapping region. Middle: Core-loss spectra (V $L_{3,2}$ and the O $K$-edges) extracted from positions indicated by color-coded arrows. Right: Horizontally projected EELS mapping showing the spectral evolution as a function of distance from the boundary. **b,** Schematic of the synergistic electro-thermal pathway: the PF effect facilitates carrier injection from oxygen-vacancy trap levels into the upper Hubbard band, boosting local conductivity and accelerating Joule heating. **c,** Schematic of the direct electronic pathway: PF-induced carrier multiplication triggers a non-thermal Mott transition via reduced electron correlation and band-gap collapse. **d,** Simulated spatial electric field profile at $t$ = 48 ns under a 3 V pulse. The bottom panel shows the electric field line profile near the top boundary. **e,** Deterministic "writing" of conduction paths via electron-beam irradiation. Left: STEM-HAADF image of Specimen 2 after line-scan irradiation (white arrow). Right: Projected core-loss EELS mapping across the irradiated region, revealing the localized generation of oxygen vacancies. **f,** SPT dynamics of Specimen 2 under microsecond pulses (0.7 V, 1000 ns). **g,** SPT dynamics of Specimen 2 under nanosecond pulses (2.5 V, 50 ns), where the nucleation and propagation are strictly confined to the pre-programmed electron-beam-irradiated line (white dashed line).

The applied field of ~1 MV m$^{-1}$ is, by itself, insufficient to trigger a purely non-thermal Mott transition[24]. Instead, our observations point to a highly cooperative electro-thermal cascade. Initially, Joule heating induces a spatially inhomogeneous temperature profile (Fig. 2d). Due to the exponential temperature-dependence of resistivity near the IMT, this thermal gradient drives a severe redistribution of the internal electric field, geometrically concentrating it at the interfaces between the thinned region and electrodes, specifically where the longitudinal boundaries meet the heat sinks (Fig. 3d). This locally enhanced field exponentially amplifies carrier concentration and promotes the Mott-driven IMT at the interfaces by the PF effect, offsetting the local thermal deficit to yield uniform nucleation across the boundaries. By transforming these boundaries into synchronized, planar heat sources, this mechanism explains the emergence of the strictly parallel domain walls that propagate inward. Upon pulse termination, the rapid thermal dissipation into the adjacent bulk regions causes the straight domain walls to curve back towards the boundary (Fig. 2c and Extended Data Fig. 6). Ultimately, while thermodynamic heating governs the global phase boundary propagation, it is the PF-mediated local field enhancement that dictates the initial spatial symmetry. Extrapolating this coupled dynamic, we project that driving the system at extreme fields (e.g., 30 V) could compress the IMT timeline below 100 ps, unlocking a pathway to deep-sub-nanosecond, field-induced switching.

Capitalizing on this defect-mediated electro-thermal synergy, we sought to deterministically program the IMT spatial pathways through nanoscale defect engineering. On a second device (Specimen 2) with identical crystallographic orientation, we utilized electron-beam irradiation to "write" a quasi-1D line of oxygen vacancies across its center (Fig. 3e, indicated by white arrow and dashed line). EELS mapping confirmed that this irradiated line exhibits a chemical identical to the natural boundaries, specifically the V $L_{3,2}$-edge red-shift and O $K$-edge $t_{2g}$ suppression (Fig. 3e, right panel).

When probed under a low-voltage regime (0.7 V, 1000 ns; Fig. 3f and Supplementary Video 5), the SPT preferentially percolates perpendicular to the current before expanding longitudinally, confirming that the oxygen-vacancy line does not merely act as a low-resistance Ohmic heat source. Strikingly, under high-field excitation (2.5 V, 50 ns; Fig. 3g and Supplementary Video 6), the system undergoes a conduction-architecture reconfiguration: a second conductive filament nucleates along the pre-programmed line, concurrent with boundary nucleation. Both channels exhibit an identical ~20 ns incubation time before expanding symmetrically.

Although the central electron-beam-irradiated line is not inherently prone to accelerated Joule heating compared to the boundaries, its presence overrides macroscopic thermal diffusion limits by locally catalyzing a PF-enhanced electro-thermal cascade. Fundamentally, this field-activated channel formation is driven by the extreme non-linearity of the PF effect. Instead of gradual thermal diffusion, the local field enhancement at the oxygen vacancy line triggers an avalanche-like formation of the conductive pathway. Consequently, this dynamically reconfigurable conduction architecture represents a tunable connectivity topology. By transitioning from the low-field to the high-field regime, the system effectively jumps from an insulating state to a fully connected metallic network along the programmed defect line. Ultimately, the ability to deterministically "write" and actively switch these topological connections at the nanoscale provides a crucial foundation for morphologically reconfigurable logic and energy-efficient neuromorphic computing.

**Strain-mediated self-organization of transient triangular domains**

Beyond anisotropy filamentary percolation, our UDF imaging reveals another striking mesoscale phenomenon: the propagating phase boundaries are not rectilinear, but self-organize into highly dynamic, periodic triangular architectures (Fig. 4a). While static or metastable triangular domains have been previously observed under macroscopic stress[36] or sustained current crowding[37], the structures reported here emerge in a strictly non-equilibrium, transient regime. During the electrically driven IMT, the lattice symmetry breaking induces a ~0.3% unit-cell volume contraction, injecting significant eigenstrain into the crystal. The dynamic competition between this elastic strain energy and the domain wall interfacial energy dictates the periodic triangular boundary architectures.

Crucially, the domain evolution is exquisitely sensitive to the pulse amplitude, reflecting a fundamental dependence on the local temperature gradient. Compared to 3 V, the steeper temperature gradient at 4.5 V confines the strain-interface competition to a narrow spatial zone, suppressing prominent triangular formation during the pulse (Fig. 4b). However, upon pulse termination, the rapid relaxation of the thermal gradient triggers a domain reconfiguration, manifested as the emergence of periodic triangular domains and continuous geometric scaling. Crystallographic analysis reveals that these transitions are deterministically guided by the intrinsic lattice symmetry: the triangular facets preferentially align with the $[110]_R$ and $[211]_R$ crystallographic directions (red and blue dashed lines) on the $(\bar{1}11)_R$ observation plane (Fig. 4d), effectively minimizing the anisotropic elastic energy density[36-38].

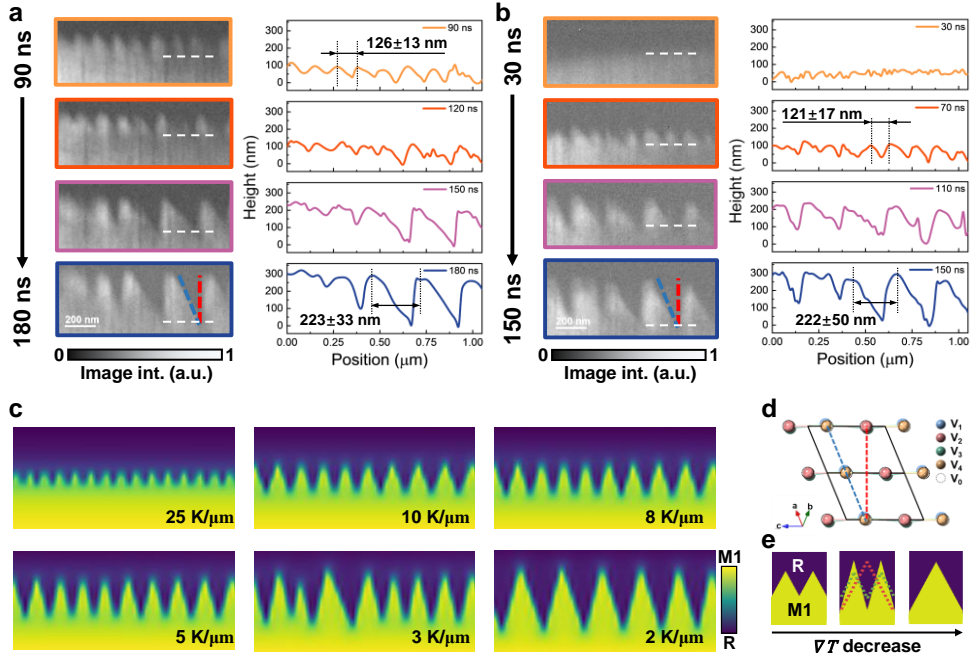

**Fig. 4. Dynamics of strain-mediated triangular domains and phase-field simulations. a,** Spatiotemporal evolution of triangular domain architectures triggered by a 3 V, 100 ns pulse. Left: Sequential UDF snapshots capturing the domain morphology at 90, 120, 150, and 180 ns. Right: Corresponding triangular domain profiles demonstrating the continues increase in domain height and progressive increase in domain periodicity. The vertical axis represents the height of the triangular domain, with the white dash line on the left panel serving as the reference. **b,** Domain evolution under a 4.5 V, 30 ns pulse. Left: Snapshots at 30, 70, 110, and 150 ns. Right: Triangular domain profiles revealing a temporally flat yet accelerated domain wall coarsening kinetics compared to **a**. **c,** Phase-field simulations of phase distributions under varying thermal gradients ($\nabla T$). A decreasing gradient promotes the transition from a rectilinear-like boundary to prominent triangular oscillations and subsequent period-doubling. **d,** Projection of the M1-phase crystal structure along the $[011]_{M1}$ direction. $V_0$ denotes the high-symmetry vanadium positions in the R phase, while $V_1$-$V_4$ denote the distinct displacement vectors of vanadium atoms during the structural transition. Red and blue dashed lines indicate the $[110]_R$ and $[211]_R$ crystallographic directions, respectively, which align with the triangular facets to minimize elastic energy. **e,** Schematic illustration of the period-doubling mechanism: as the temperature gradient relaxes, triangular domains increase in amplitude and undergo stochastic merging, leading to a discrete doubling of the spatial period.

As the temperature gradient relaxes and continuous morphological distortion escalates, we observe a remarkable period-doubling event occurring within a mere 50 ns post-pulse. During this transition, adjacent triangular domains abruptly coalesce, annihilating their intermediate vertices to yield a coarsened doubly-spaced periodic architecture. This gradient-dependent structural evolution is rigorously corroborated by our temporal phase-field simulations (Fig. 4c). By dynamically relaxing the temperature gradient from 25 K µm$^{-1}$ to 2 K µm$^{-1}$, the numerical model faithfully reproduces the experimental trajectory: a repetitive sequence of continuous amplitude growth punctuated by discrete, stepwise period-doubling bifurcations (See Extended Data Fig. 8 and Supplementary Video 7 for detail).

Integrating these experimental and simulated results with our energetic analysis, the entire non-equilibrium dynamic evolution can be unified into a cohesive physical picture (Fig. 4e): Initially, the highly anisotropic elastic and interfacial energies rigidly pin the domain walls along specific, energetically favorable crystallographic planes. As the thermal gradient relaxes, the domains are forced to elongate, severely distorting these pinned walls and accumulating a significant elastic-interfacial energy penalty. To dissipate this localized energy peak, the system undergoes a rapid morphological merger, annihilating adjacent vertices to snap the domain walls back to their pristine, pinned crystallographic orientations, which simultaneously doubles the spatial period. This elegant mechanism minimizes the global free energy while driving the macroscopic phase progression. The coexistence of continuous geometric scaling and discrete structural reconfiguration within a single device establishes a unified physical paradigm for non-equilibrium phase transitions, demonstrating that the macroscopic IMT is governed by a sophisticated interplay between continuous morphological distortion and quantized configurational resets.

**Conclusion**

In summary, leveraging the high-resolution, real-space, and time-resolved imaging capabilities of our custom-designed E-UTEM, we have unraveled the multi-scale electro-thermo-mechanical dynamics governing the IMT in $VO_2$. Our findings demonstrate that the synergistic interplay between intrinsic electro-thermal coupling and the PF effect induces a dynamic redistribution of the internal electric field, triggering localized, field-driven transitions. This mechanism serves a dual role in redefining the operational landscape of $VO_2$-based devices. First, the PF effect significantly accelerates the IMT kinetics, providing a deterministic physical pathway to compress switching speeds into the sub-100 ps regime. Second, capitalizing on the extreme nonlinearity of the PF effect, we have achieved dynamically reconfigurable connectivity topologies within a single device. By precisely steering the spatial nucleation of conductive pathways via defect engineering and field modulation, we transition from fixed-circuitry limitations to fluidic, field-responsive architectures.

Beyond field-steered initiation, this electronic switching further sets the stage for a complex mesoscale structural evolution. We find that the resultant phase progression is not merely a passive thermal expansion but a highly organized process governed by the coupling of thermal and elastic energies. In this regime, the continuous morphological distortion of phase boundaries is periodically resolved through discrete, period-doubling configurational resets, effectively minimizing the global energy penalty during non-equilibrium progression.

By unifying the field-driven electronic initiation and strain-mediated structural evolution, our findings establish a comprehensive physical paradigm for non-equilibrium phase transitions. This demonstrates that the macroscopic IMT is governed by a sophisticated interplay between quantized electronic switching and quantized structural reconfiguration. Ultimately, the ability to deterministically "write" and actively steer these topological connections at the nanoscale provides a robust hardware foundation for morphologically reconfigurable logic and high-speed, energy-efficient neuromorphic computing architectures, paving the way for the next generation of adaptive electronic systems.

**Method**

**Device Fabrication: Single-Crystalline VO$_2$ Nano-Device**

Single-crystalline VO$_2$ particles were synthesized via high-temperature recrystallization process. Amorphous $VO_2$ precursors (Tangchuan Sci-tech Co., Ltd) were encapsulated in an evacuated quartz tube and annealed at 800 °C for 24 hour. To ensure high crystallinity, the sample was cooled to room temperature at a controlled rate of 5 °C min$^{-1}$, yielding single-crystalline $VO_2$ particles with typical dimensions of 20 to 30 μm.

The $VO_2$ two-terminal devices, designed for compatibility with the *in-situ* electrical TEM holder (Protochips, Fusion AX), were fabricated using a focused ion beam (FIB) system (Helios NanoLab 600i DualBeam, FEI Inc.), as illustrated in the Extended Data Fig. 1b. First, pre-existing electrodes on the electrical chip (Protochips, E-FEF01-A2) were bridged using a silver wire, which was subsequently severed at the midpoint to create a precise 6 μm gap. Second, a single-crystalline $VO_2$ specimen (approximately 8 μm in width) was extracted from a micro-particle and positioned across this gap. Then, Pt electrodes were deposited via FIB-induced deposition onto the $VO_2$ surface, establishing robust electrical contact with the silver leads. The resulting electrode-to-electrode spacing was approximately 6 μm. Finally, to enable electron transparency for UDF imaging, the central 3.1 μm region of the suspended $VO_2$ device was thinned to a target thickness of approximately 180 nm.

**Ultrafast Pulse Generation and Synchronization (E-UTEM)**

The electric-pump ultrafast transmission electron microscopy (E-UTEM) was implemented on a commercial JEOL JEM-2100Plus platform at the Synergetic Extreme Condition User Facility (SECUF). A comprehensive schematic of the E-UTEM architecture is provided in Extended Data Fig. 1a. To generate ultrafast electron pulses, a femtosecond laser system (Spirit 1040-4, Spectra-Physics; 1040 nm, 300 fs, 10 kHz) was integrated with the electron gun, where it irradiated the $LaB_6$ filament to trigger photoemission. The synchronization of the electric pump and the probe electron beam followed a master-trigger protocol: (1) A fraction of the Spirit laser beam was directed onto a photodiode to generate an initial electrical pulse. (2) This pulse was fed into a digital delay generator (DG645, Standford Research Systems) for external triggering. The DG645 served as the central timing controller, outputting two synchronized channels with a jitter of 25 ps. (3) One channel externally triggered the nanosecond laser (Stele-A-1, 355 nm, 10 ns duration), while the second channel delivered the electrical excitation to the *in-situ* specimen holder. The DG645 generated square wave pulses with a 2 ns rise time and adjustable amplitudes (0.5-5 V). The temporal evolution of the specimen was mapped by adjusting the relative delay between the electron probe and the electric pump via the DG645. Ultrafast imaging and diffraction patterns were captured using a high-speed CMOS camera (Gatan OneView).

To ensure signal integrity and enable real-time monitoring of the device response, the electrical pump signal was divided into two parallel circuits prior to entering the *in-situ* holder (Extended Data Fig. 1d). The primary circuit consisted of the $VO_2$ device in series with Channel 1 (CH1) of a wide-bandwidth oscilloscope (50 Ω input impedance), recording the transmitted voltage ($V_t$). The parallel bypass was connected to Channel 2 (CH2, 50 Ω input impedance) to monitor the incident voltage ($V_i$) and maintain impedance matching. Accounting for signal reflections within the system, the incident voltage $V_i$ relates to the source output $V_0$ as:

$$V_i = \frac{R+50}{R+75} V_0 \qquad (2)$$

The transient resistance of the $VO_2$ device ($R$) was calculated in real-time based on the voltage division:

$$R = 50 \frac{V_0}{V_t} - 75 \quad \text{or} \quad R = 50 \frac{V_i}{V_t} - 50 \qquad (3)$$

**Electron Energy-Loss Spectroscopy (EELS)**

EELS was performed using a Gatan DualEELS (GIF Continuum K3 1069) system integrated with a JEOL NEOARM 200F TEM (operating at 200 kV) at the SECUF. Spectra were acquired in DualEELS mode with an energy dispersion of 0.09 eV per channel, allowing for the simultaneous capture of low-loss and core-loss regions. To ensure high precision in chemical shift analysis, all spectra were energy-calibrated using the simultaneously acquired zero-loss peak (ZLP). The energy resolution, defined by the full-width at half-maximum (FWHM) of the ZLP, was approximately 0.8 eV under the experimental conditions. Background subtraction and plural scattering deconvolution were performed using the Gatan Microscopy Suite to extract the intrinsic fine structures of the V-$L_{2,3}$ and O-$K$ edges.

**Electron-Beam Irradiation and Defect Engineering**

Controlled electron-beam irradiation was performed using the JEOL NEOARM 200F TEM operating at an acceleration voltage of 200 kV. To ensure precise spatial patterning of oxygen vacancies, the microscope was operated in scanning (STEM) mode with a probe current of 160 pA. Specific conduction pathways were defined using a line-scan protocol with a step size of 2.5 nm and a pixel dwell time of 2 s. This configuration yields a cumulative linear electron dose of approximately $4\times10^6$ e$^-$ Å$^{-2}$. The beam was accurately positioned using the microscope's scan controller to target specific device regions. This localized high-dose irradiation was designed to trigger oxygen displacement and the subsequent formation of vacancy-rich filaments, thereby modulating the local electronic landscape and Poole-Frenkel response.

**Electro-Thermal Finite Element Simulation**

To investigate the transient temperature and field distributions, three-dimensional finite element simulations were conducted using COMSOL Multiphysics 6.0 (Electromagnetic Heating module). The 3D model was constructed to replicate the actual specimen geometry. The thickness profile of the central electron-transparent region was calibrated using EELS log-ratio thickness mapping (Extended Data Fig. 3). To accurately represent the heat-sinking effect and current injection from the bulk of the device, rectangular blocks (1 μm × 1 μm × 5 μm) were integrated on both sides of the thinned membrane.

The electrical excitation was implemented via the 'Electric Potential' boundary condition for pulsed-voltage simulations and the 'Normal Current Density' condition for constant-current analysis. Thermal dissipation from the left and right boundaries was modeled through a prescribed heat flux $q = -35[W/m/K] \times \frac{(T-293.15[K])}{1[\mu m]}$ to account for conductive heat dissipation through the electrodes, where the coefficient $35[W/m/K]$ is half of the thermal conductivity of Pt. The remaining boundaries were set as thermally insulated. The base temperature fixed at 293 K.

The electrical conductivity of the M1 phase $\sigma_{M1}$ was estimated using the Poole–Frenkel expression[18,22] and the *I–V* curve in Extended Data Fig. 1c:

$$\sigma_{M1} = 3.5 \times 10^4 \times e^{-\frac{608}{T}} \text{ S m}^{-1} \qquad (4)$$

While the electrical conductivity of the R phase $\sigma_R$ was set to $1 \times 10^5$ S m$^{-1}$. The analytical expression for the total electrical conductivity $\sigma$ is given by:

$$\sigma = \sigma_{M1} \times \left(1 - \frac{1}{1+e^{-a*(T-T_c)}}\right) + \sigma_R \times \frac{1}{1+e^{-a*(T-T_c)}} \qquad (5)$$

The parameter $a$, which determines the steepness of the transition, was set to 6. The parameter $T_c$, defining the midpoint of the transition temperature, was set to 339 K.

The heat capacities of the M1 and R phases were set to $c_{low}$ = 656 J kg$^{-1}$ K$^{-1}$ and $c_{high}$ = 780 J kg$^{-1}$ K$^{-1}$, respectively. The latent heat of the phase transition, $L$ = 51.8 kJ kg$^{-1}$ [39], was incorporated into the heat capacity within a $\omega_{trans}$ = 2 K temperature window around the transition point $T_{trans}$ = 338 K:

$$c_p = \frac{2 \cdot L}{\omega_{trans}} \cdot \sin^2\left(\frac{(T-T_{trans}) \cdot \pi}{\omega_{trans}}\right) + c_{low} + (c_{high} - c_{low}) \cdot \sin^2\left(\frac{(T-T_{trans}) \cdot \pi}{2 \cdot \omega_{trans}}\right) \quad (6)$$

The thermal conductivity was set to 6 W m$^{-1}$ K$^{-1}$ [23] and the density to 4600 kg m$^{-3}$ [39].

**Phase-Field Simulations**

To capture the mesoscale structural evolution and the period-doubling dynamics, we employed a phase-field model based on the Ginzburg-Landau theory. Three-dimensional phase-field simulations were carried out by solving the Cahn-Allen dynamics equation[38]:

$$\frac{\partial \phi}{\partial t} = -L \frac{\delta F}{\delta \phi} \quad (7)$$

The phase state is described by a non-conserved order parameter $\phi$, where $\phi$ = 1 denotes the M1 phase and $\phi$ = 0 denotes the R phase. $L$ is a coefficient related to the domain wall mobility, and $F$ denotes the total free energy of the system[36]:

$$F(\phi) = \int \left[f(\phi) + \frac{\beta^2}{2}|\nabla\phi|^2 + \frac{1}{2}C_{ijkl}(\varepsilon_{ij} - \varepsilon_{ij}^T)(\varepsilon_{kl} - \varepsilon_{kl}^T)\right] d\Omega \quad (8)$$

which includes contributions from the thermodynamic energy density $f(\phi)$, the interfacial energy density $\frac{\beta^2}{2}|\nabla\phi|^2$, and the strain energy density $\frac{1}{2}C_{ijkl}(\varepsilon_{ij} - \varepsilon_{ij}^T)(\varepsilon_{kl} - \varepsilon_{kl}^T)$. The thermodynamic energy density is formulated as a temperature-dependent double-well potential[40]:

$$f(\phi) = \frac{\phi^4}{4} - \left(\frac{1}{2} - \frac{m}{3}\right)\phi^3 + \left(\frac{1}{4} - \frac{2}{m}\right)\phi^2 \quad (9)$$

where m is a temperature-dependent parameter governing the deviation from phase equilibrium, implemented following Kobayashi's prescription $m = \frac{\alpha}{\pi}\tan^{-1}(\gamma(T_e - T))$. $\alpha$ and $\gamma$ are phenomenological parameters, $T_e$ is the phase transition temperature. The interfacial energy density depends on the gradient of the order parameter $\phi$, where the gradient coefficient $\beta$ is calibrated to yield a surface energy of approximately 25 mJ m$^{-2}$ [41]. The strain energy density is calculated based on microelasticity theory, in which the elastic modulus tensor is adopted from the literature[42], which contains six independent parameters: $C_{1111}$=342 GPa, $C_{1122}$=253 GPa, $C_{1133}$=175 GPa, $C_{3333}$=434 GPa, $C_{1313}$=137 GPa, $C_{1212}$=127 GPa. All the parameters were scaled proportionally based on the measured value of $C_{3333} \approx 140$ GPa near room temperature[43]. $\varepsilon$ is the total strain. Eigen strain tensor $\varepsilon^T$ associated with the SPT from M1 to R is in the form: $\varepsilon_{xx}^T = 0.006(1-\phi)$, $\varepsilon_{yy}^T = 0.001(1-\phi)$, $\varepsilon_{zz}^T = -0.01(1-\phi)$. $x, y$ and $z$ correspond to the directions of the a-, b-, and c-axes of the R phase, respectively. The temporal evolution of the phase distribution and the stress/strain field follows the coupled Cahn-Allen dynamics equation and the mechanical equilibrium equations, which are solved in COMSOL Multiphysics 6.0 (Solid Mechanics module and Weak Form PDE module).

**Data availability**

All the data are available from the corresponding author upon reasonable request.

**Acknowledgements:** This work was supported by the National Natural Science Foundation of China, Grant No. U22A6005; the National Key R & D Program of China, Grant Nos. 2021YFA13011502, 2024YFA1611303 ,2024YFA1408701 and 2024YFA1408403; Yin Shi was supported by the start-up grant from the Institute of Physics, Chinese Academy of Sciences; This work was carried out at the Synergetic Extreme Condition User Facility (SECUF, https://cstr.cn/31123.02.SECUF ).

**Author contributions:** Jun Li and Jianqi Li conceived the study and supervised the experimental part of the project. Siyuan Huang carried out the experiments, numerical simulations and analyzed the data. Shuaishuai Sun, Wentao Wang and Huanfang Tian carried out the design of the experimental setup. Yin Shi supervised the numerical simulations. Chunhui Zhu provided the samples. Huaixin Yang and all other authors participated in discussions. Siyuan Huang, Jun Li and Yin Shi wrote the manuscript with input from all other authors.

**Competing interests:** The authors declare no competing interests.

**Correspondence and requests for materials** should be addressed to Jun Li or Jianqi Li.

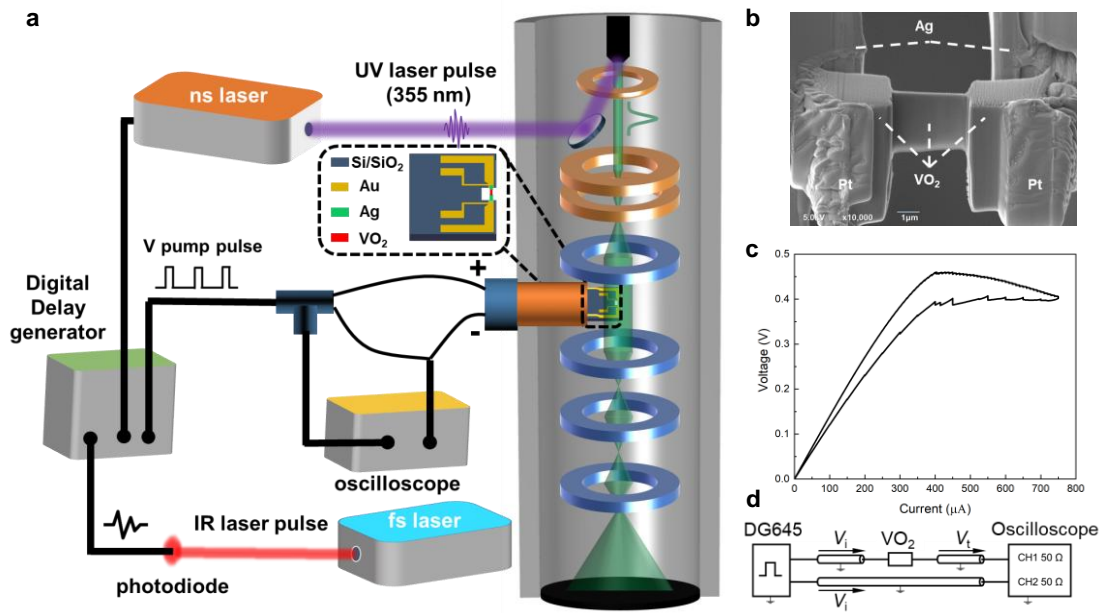

**Extended Data Fig. 1.** *Operando* **experimental setup and electrical measurement. a,** Extended Schematic of the E-UTEM setup. **b,** SEM image of the fabricated $VO_2$ specimen across the slit of an *in-situ* electrical chip. **c,** Voltage–current loops of the $VO_2$ two-terminal device. The current ramp rate is set to 10 µA s$^{-1}$ for the increasing sweep and 50 µA s$^{-1}$ for the decreasing sweep. **d,** Schematic of the high-speed voltage measurement circuit.

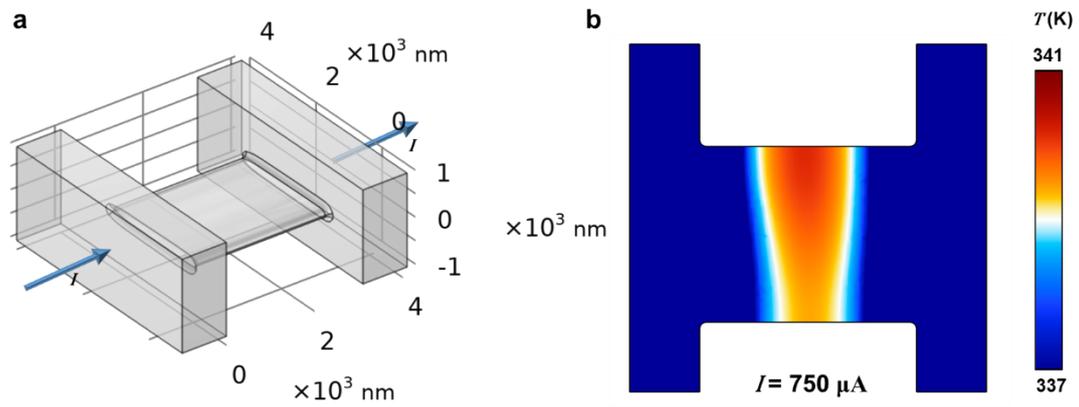

**Extended Data Fig. 2. Electro-thermal simulation. a,** Electro-thermal simulation model. The current flows in from the left boundary and out from the right boundary. **b,** Spatial temperature profile obtained from the electro-thermal simulation. The current was ramped up at a rate of 10 µA s$^{-1}$ to a final value of 750 µA. The image corresponds to the state at t = 75 s, i.e., I = 750 µA.

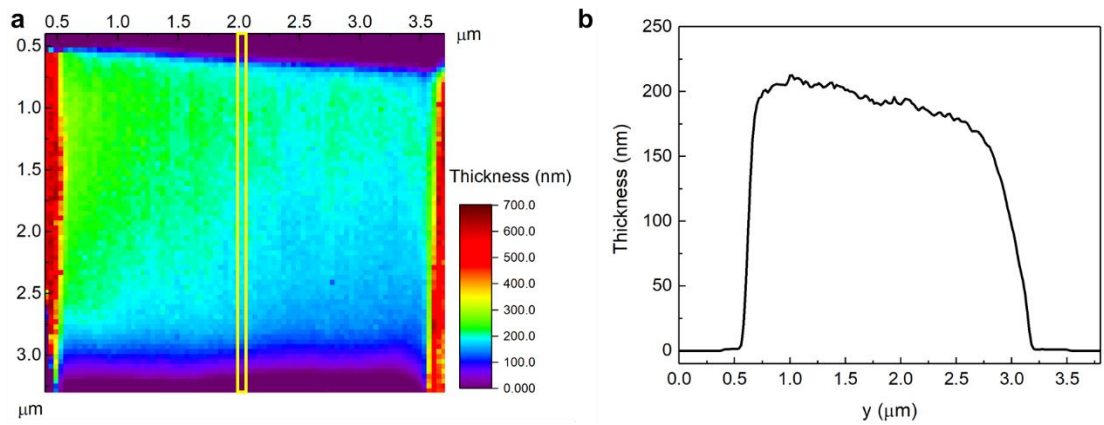

**Extended Data Fig. 3. Thickness measurement. a,** Thickness mapping of the sample obtained from EELS analysis. **b,** Thickness profile across the section marked by the yellow box in **a**. The thickest region of the sample is approximately 200 nm, with an average thickness of about 180 nm.

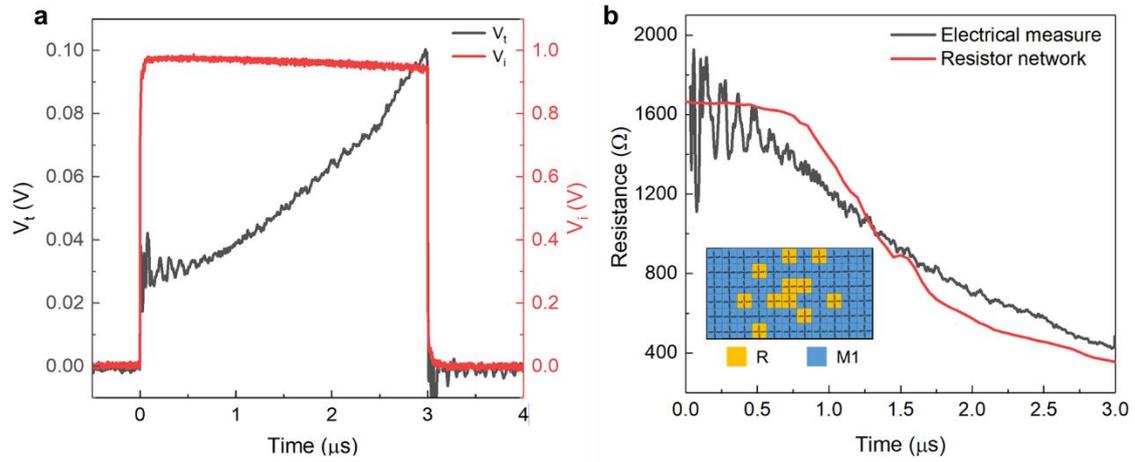

**Extended Data Fig. 4. High-speed electrical measurement and resistance response.**
**a,** Temporal evolution of the incident voltage $V_i$ and transmitted voltage $V_t$ during the phase transition driven by a 1 V, 3000 ns pulse. The significant rise in $V_t$ after ~600 ns indicates the onset of the phase transition following an incubation period, corresponding to a decrease in sample resistance. **b,** The resistance change calculated through electrical measurements according to Equation (3) (black line) and derived from the resistor network model (red line). The inset is a schematic of the resistor network.

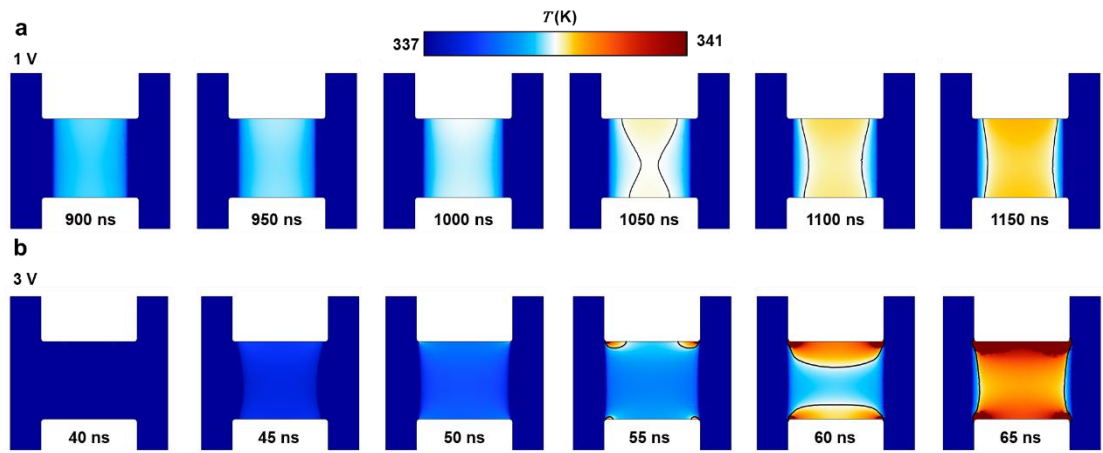

**Extended Data Fig. 5. Electro-thermal simulation under pulse excitation. a,** Simulated spatial temperature profile at different time for the 1 V pulse. **b,** Simulated spatial temperature profile at different time for the 3 V pulse. The black curve denotes the 339 K isotherm.

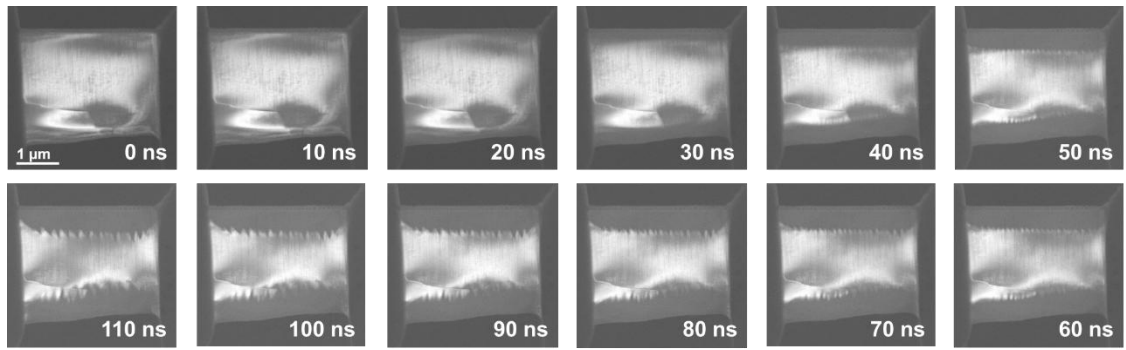

**Extended Data Fig. 6. Ultrafast DF images of the transient domain dynamics under a 4.5 V, 30 ns pulse.**

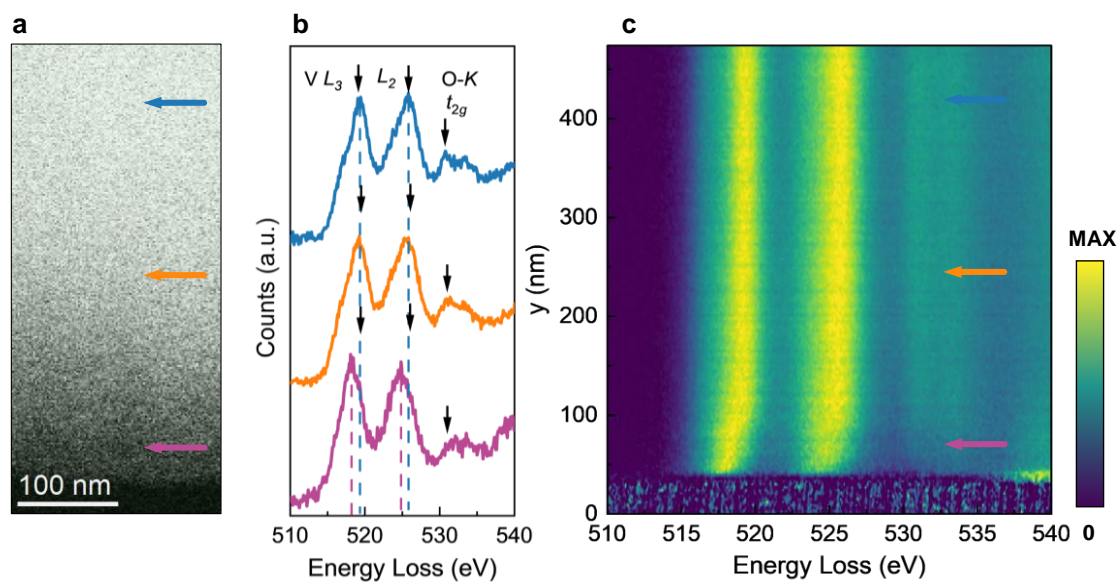

**Extended Data Fig. 7. EELS near the bottom boundary of the sample. a,** STEM-HAADF image of the EELS mapping region. **b,** Core-loss spectra (V $L_{3,2}$ and the O $K$-edges) extracted from positions indicated by color-coded arrows in **a**. **c,** Horizontally projected EELS mapping showing the spectral evolution as a function of distance from the boundary.

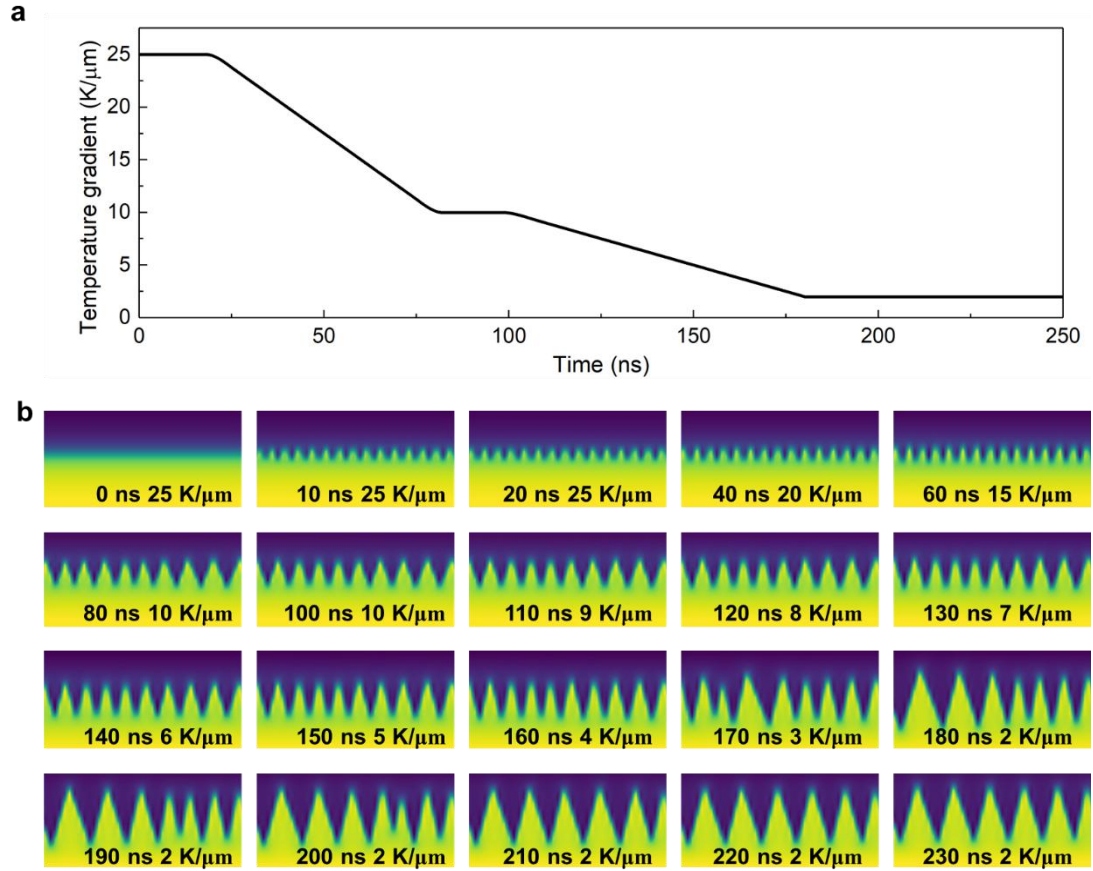

**Extended Data Fig. 8. Phase-field simulations of phase distributions under varying thermal gradients (∇*T*). a,** Temporal evolution of the temperature gradient in the phase-field simulations. **b,** Phase distributions obtained from phase-field simulations under different time and temperature gradient conditions.

**Extended Data Table 1. Time constants of the M1 volume fraction and normalized resistance variation under different voltages.**

| Voltage | | Time constant / ns | Phase transition fraction or Resistance change ratio |
|---|---|---|---|
| 1 V | M1 volume fraction | 560 ± 29 | 0.70 |
| | $R_{//}/R_0$ | 371 ± 11 | 0.77 |
| | $R_{\perp}/R_0$ | 285 ± 11 | 0.90 |
| 3 V | M1 volume fraction | 21.1 ± 0.9 | 0.46 |
| | $R_{//}/R_0$ | 10.8 ± 0.4 | 0.76 |
| | $R_{\perp}/R_0$ | 22.5 ± 0.7 | 0.50 |
| 4.5 V | M1 volume fraction | 8.9 ± 0.3 | 0.37 |
| | $R_{//}/R_0$ | 5.8 ± 0.2 | 0.71 |
| | $R_{\perp}/R_0$ | 10.9 ± 0.5 | 0.40 |